# Giant Topological Hall Effect Across Wide Temperature in Pt/NiCo$_2$O$_4$ Heterostructure

Bharat Giri,[1] Ahsan Ullah,[1] Jing Li,[1] Bjorn Josteinsson,[2] Zhewen Xu,[2] Suvechhya Lamichhane,[1] Adam Erickson,[3] Arjun Subedi,[1] Peter A Dowben,[1] Gabriel Puebla Hellmann,[2] Abdelghani Laraoui,[1,3,4] Sy-Hwang Liou,[1,4] Xiaoshan Xu[1,4]*

[1]Department of Physics and Astronomy, University of Nebraska, Lincoln, Nebraska 68588, USA

[2]QZabre LLC, Zurich 8050, Switzerland

[3]Department of Mechanical & Material Engineering, University of Nebraska, Lincoln, Nebraska 68588, USA

[4]Nebraska Center for Materials and Nanoscience, University of Nebraska-Lincoln, Nebraska 68588, USA

*Email: xiaoshan.xu@unl.edu

**Abstract**

Topological Hall effect (THE), a quantum phenomenon arising from emergent magnetic field generated by topological spin texture, is a key method for detecting non-coplanar spin structures like skyrmions in magnetic materials. Here, we investigate a bilayer structure of Pt and conducting ferrimagnet NiCo$_2$O$_4$ (NCO) of perpendicular magnetic anisotropy and demonstrate giant THE across a temperature range 2–350 K. The absence of THE in single-layer Pt and NCO, as well as in Pt/Cu/NCO, suggests its interfacial origin. The maximum THE occurring just before the NCO coercive field indicates its connection to magnetic nucleation centers, which are topologically equivalent to skyrmions. The large normalized THE, based on the emergent-field model, points to a high population density of small nucleation centers. This aligns with the unresolvable domain structures during magnetization reversal, even though clear domain structures are detected after zero-field cooling. These results establish heavy metal/NCO as a promising system for exploring topological spin structures.

Topological spin textures are spatial arrangement of spins that get extra stability from their non-trivial topology,[1] with skyrmions as good examples that carry quantized magnetic flux. The distinct interactions between topological spin textures and itinerant electrons lead to effective control by electric field, which makes them compelling for next-generation memory and logical spintronics devices. [1–3]

Topological Hall effect (THE) is a quantum phenomenon resulting from the exchange interactions that align the spins of itinerant electrons with local spins in a non-coplanar spin texture. The effect can be described using an emergent magnetic field $\vec{B}_e$,[4] which deflects itinerant electrons via Lorent force and causes of the transverse signal of THE. $\vec{B}_e$ depends on how rapid local spin changes in a non-coplanar fashion, so it is expected to be zero when magnetization is saturated but maximized near coercive field in the magnetization reversal process. These features of THE have been a key tool to probe topological spin textures.[5,6]

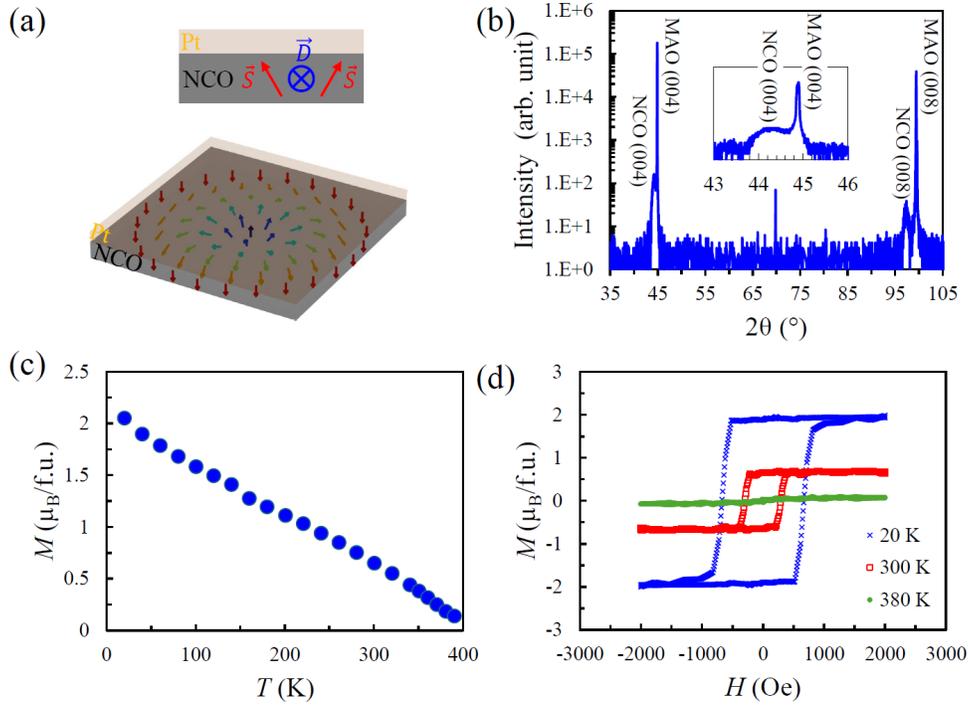

**Figure 1.** (a) Schematic diagram of spin canting by interfacial DMI ($\vec{D}$) at the Pt-NCO interface (top) and the resulting topological spin structure (bottom). Black arrows indicate the magnetic moments at the interface in NCO. (b) θ-2θ scan of an NCO (12 nm) / MAO (001) film. The inset is a closeup view of the NCO (004) peak. (c) Temperature dependence of magnetization of Pt (3 nm) / NCO (15 nm) / MAO (001) film measured in an out-of-plane magnetic. (d) Field dependence of magnetization (*M-H*) loops at various temperatures of the same sample in (c).

In this work, we explore topological spin texture in bilayer system Pt/NiCo$_2$O$_4$ using THE. NiCo$_2$O$_4$ (NCO) is a recently discovered conductive ferrimagnet with Curie temperature $T_C \approx 420$ K [7,8]. Epitaxial NCO films grown on MgAl$_2$O$_4$ or MAO (001) exhibit strong perpendicular magnetic anisotropy or PMA of 0.2 MJ m$^{-3}$ at room temperature.[7,9] As illustrated in **Figure 1(a)**, the heavy metal Pt on top of NCO may introduce interfacial Dzyaloshinskii-Moriya interaction (DMI) $\vec{D}$ [10–12] and promote topological spin textures. For magnetic materials with PMA, the nucleation centers generated during the magnetization reversal process are topologically equivalent to Neél's type of skyrmions,[13] *i.e.*, spin texture in which the magnetization gradually rotates from the perpendicular direction in the center to the opposite direction at the perimeter (**Figure 1(a)**). Interfacial DMI is expected to reduce the critical size of these nucleation centers and increase their population causing substantial $\vec{B}_e$ and THE.

Here, we demonstrate a strong THE in the Pt/NCO thin films, as indicated by peaks/dips of Hall signal near coercivity of magnetization reversal in addition to the nearly-square-shaped anomalous Hall effect (AHE) signal from NCO, across the temperature range studied 2–350 K. The appearance of THE maximum before coercivity suggests its relationship to the magnetic nucleation centers which are topologically equivalent to skyrmions. The maximum THE, in both Hall resistivity and normalized Hall resistivity, exceeds most known reported values in the multilayer systems. Using the emergent-field model, we estimate a population density of nucleation center up to $7 \times 10^3$ μm$^{-2}$ which corresponds to sizes of about 10 nm near the coercivity of magnetization reversal. The small size of the nucleation centers is consistent with unresolvable magnetic domain structures during magnetization reversal by scanning probe microscopies, even though distinct magnetic domain structures are observed after zero-field cooling.

First, we show PMA in NCO film epitaxially grown on a MAO (001) substrate. X-ray diffraction (XRD) of a typical NCO/MAO (001) film is shown in **Figure 1(b)**. The inset is a close-up view near the NCO (004) peak, suggesting no other crystalline phases. The out-of-plane (OOP) strain of the NCO film calculated from the XRD peak positions is ≈ 1.25%, consistent with previous results.[9] This strain is expected to generate PMA.[9] As shown in **Figure 1(c)**, temperature dependence of the magnetization (*M-T*) measured during warming in 5 kOe OOP magnetic field shows that the magnetic transition temperature is 395 K. The magnetization at 20 K is 2.1 μ$_B$ /f.u. which agrees with the previously reported values[7] for NCO films grown in optimal conditions.[14] Nearly-square-shaped magnetization-field (*M-H*) loops at 20 K and 300 K (**Figure 1(d)**) are consistent with PMA of the NCO films.[9]

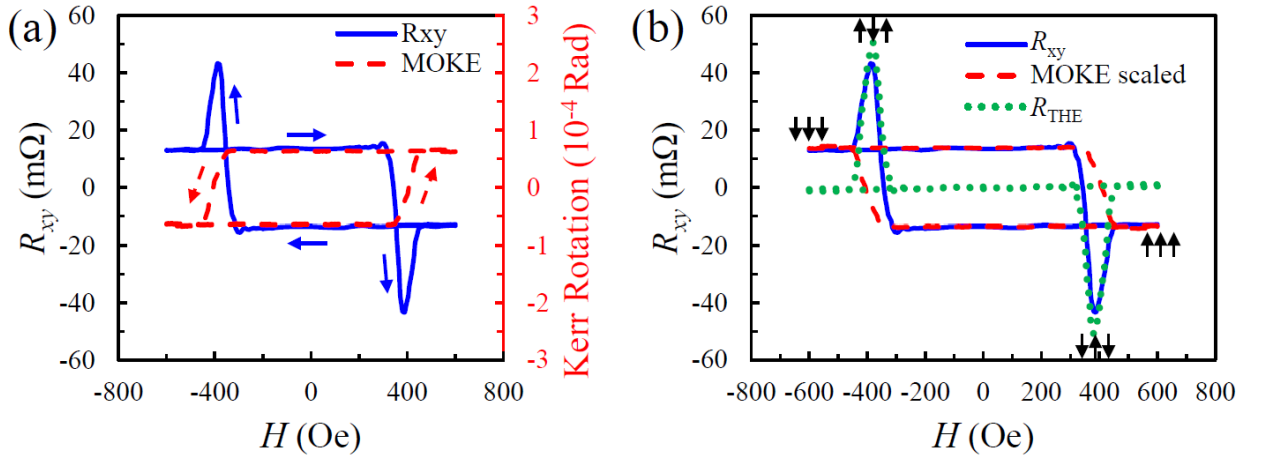

**Figure 2.** (a) Hall and MOKE measured simultaneously on a Pt (3 nm) / NCO (15 nm) / MAO (001) film at 295 K. Arrows show the history of *H* change. (b) Extraction of the topological Hall effect. The three-arrow groups indicate the magnetic states. ↑↑↑ and ↓↓↓ indicate saturated state. ↓↑↓ and ↑↓↑ indicate nucleation centers.

Giant THE is observed in a Pt (3 nm)/NCO (15 nm) sample (see thickness measurement in **Figure S1**, Supporting Information). **Figure 2(a)** shows room temperature Hall measurement on the Pt (3 nm)/NCO (15 nm) film with an OOP magnetic field; the arrows indicate the history of the applied field. The Hall signal includes mainly a nearly-square-shaped AHE loop, which is consistent with the magnetization hysteresis loop measured by the magneto-optical Kerr effect (MOKE) simultaneously on the same Hall bar, as also shown in **Figure 2(a)**. In addition, the Hall effect curve also exhibits a peak and a dip near the coercive field, suggesting THE. Moreover, measurements on NCO/MAO films of similar thickness without the Pt layer (**Figure S2**, Supporting Information), Pt/Cu/NCO/MAO films with a Cu layer inserted between Pt and NCO (**Figure S3**), and Pt layer alone on non-magnetic insulating substrates[15] do not show similar behavior, indicating that the observed THE in **Figure 2(a)** comes from the Pt/NCO interface.

We quantify the THE signal by subtracting AHE from the total Hall resistance $R_{xy}$. The contributions to Hall resistance can be decomposed into $R_{xy} = R_{OHE} + R_{AHE} + R_{THE}$, where $R_{OHE}$ is the ordinary Hall resistance which is negligible in this low-field conditions, $R_{AHE}$ is the anomalous Hall resistance which is proportional to magnetization, and $R_{THE}$ is the topological Hall resistance. Hence, assuming that the MOKE signal is proportional to the magnetization and $R_{AHE}$, $R_{THE}$ can be found by subtracting a scaled MOKE signal from $R_{xy}$. We note that the Hall and MOKE signals were measured simultaneously in the same magnetic field, which precludes any offset of magnetic field between the two measurements due to the field calibration difference. **Figure 2(b)** shows the individual contributions of THE and AHE. Interestingly, the THE peak and dip occur (at field defined as $H_T$) before the magnetic field reaches coercivity $H_C$.

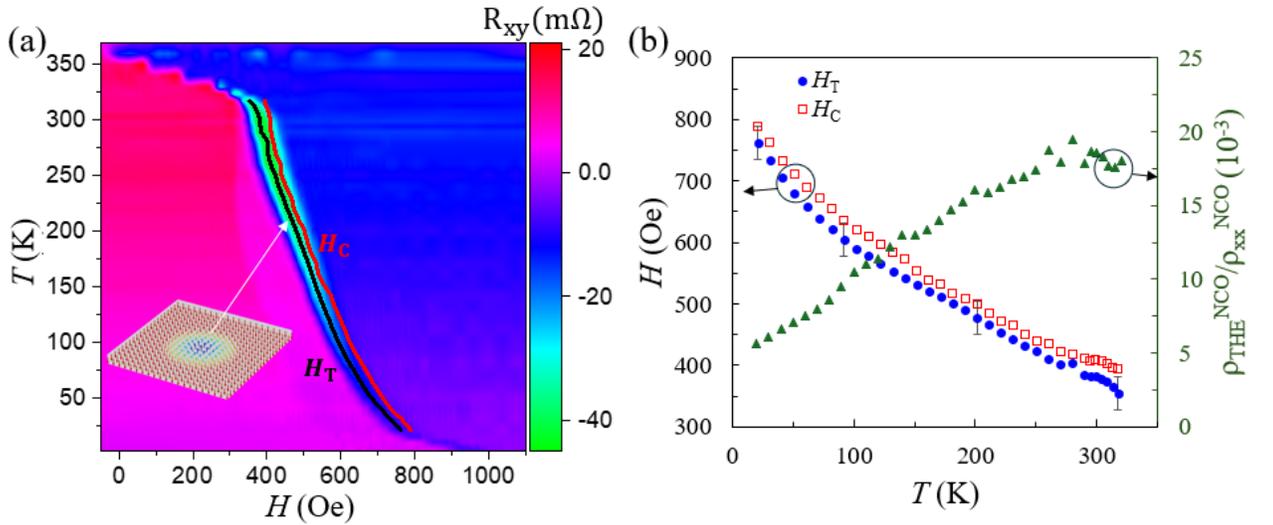

**Figure 3.** (a) Contour map of $R_{xy}$ that represents hysteresis loops with increasing magnetic field at various temperatures. The red and black curves are the coercive field ($H_C$) from MOKE and the magnetic field ($H_T$) at which dips in $R_{xy}$ are observed respectively. The inset is an illustration of magnetic nucleation center where spins are along the perpendicular direction at the perimeter and gradually rotate to the opposite direction in the center. (b) Temperature dependence of the $H_C$ and $H_T$ (error bars represent FWHM of THE peaks) and the normalized topological Hall resistivity.

Half of the hysteresis loop (increasing magnetic field) of $R_{xy}$ is plotted for the temperature range 2–370 K in **Figure 3(a)**. THE is observed in the measured temperature range 2–350 K (see **Figure S4**, Supporting Information), with MOKE measurements at different temperatures are shown in **Figure S5**. The temperature dependent $H_T$ and $H_C$ are plotted in both **Figures 3(a and b)**. The error bars of $H_T$ are the full width at half maximum (FWHM) of the THE dips. As temperature increases, both $H_T$ and $H_C$ decrease; $H_T$ consistently follows and remains smaller than $H_C$. The THE resistivity, normalized with longitudinal resistivity for the NCO film are also plotted in **Figure 3(b)**. A contour map of normalized topological Hall resistivity of the complete cycle is shown in **Figure S6**, Supporting Information.

From **Figure 3(b)**, it is clear that $H_T < H_C$ holds for all temperatures. This consistent relation suggests that THE is connected with the nucleation process of the magnetization reversal.[16] For materials of PMA, nucleation centers have magnetization points perpendicular to the film plane at the perimeter but along the opposite direction in the center. This configuration is topologically equivalent to skyrmions,[16–18] as illustrated in **Figure 3(a)** inset.

We therefore investigate the nucleation process during the magnetization reversal at room temperature by measuring magnetic domain patterns using scanning probe microscopies.

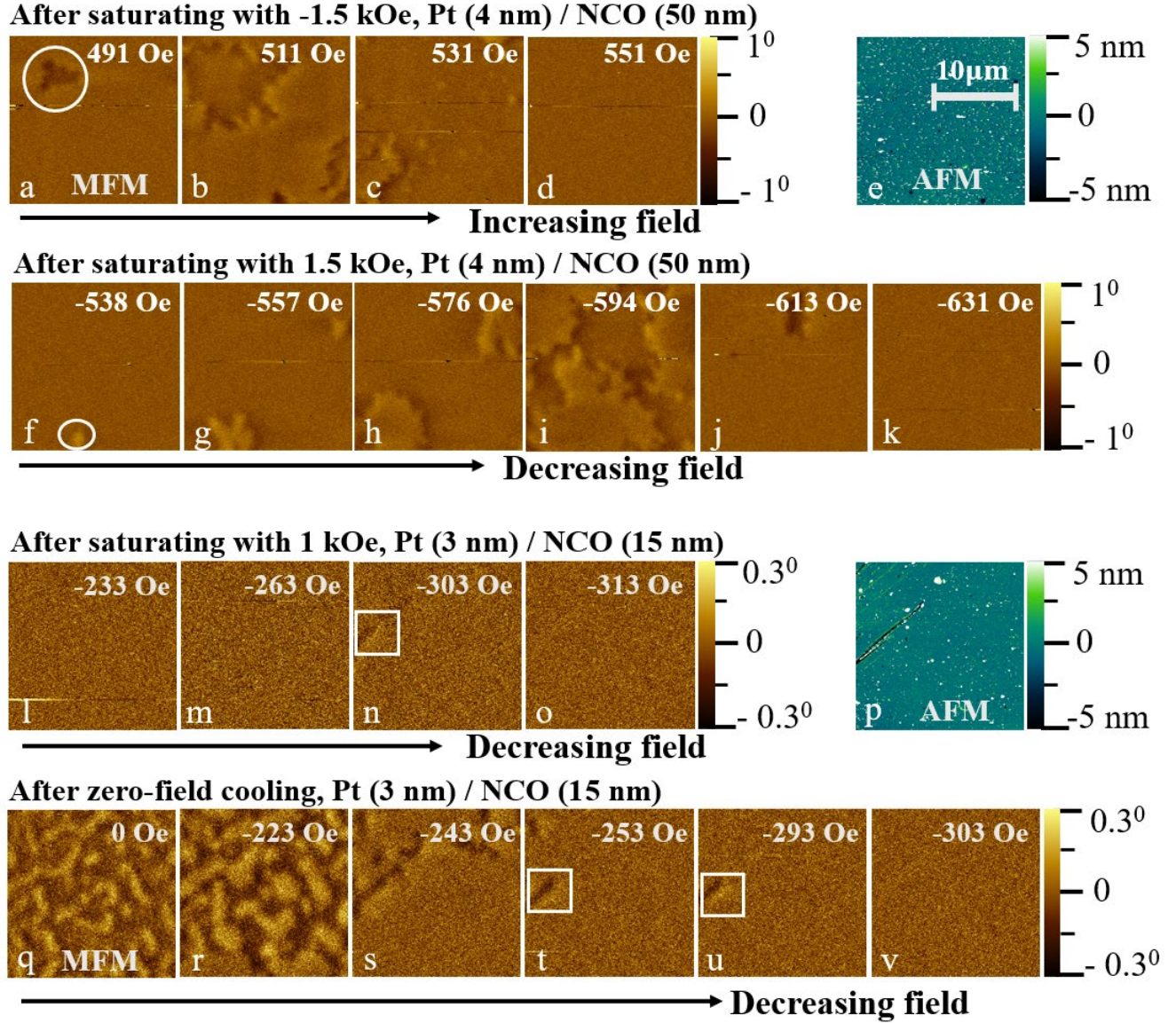

**Figure 4.** MFM images (20 × 20 μm²) of the Pt (4 nm) / NCO (50 nm) / MAO (001) film (a-d & f-k) and the Pt (3 nm) / NCO (15 nm) /MAO (001) film (l-o & q-v) at room temperature. (a-d) The Pt (4 nm) / NCO (50 nm) film with increasing field after saturating in a -1.5 kOe field. The circles in (a & f) indicate nucleation centers. (e) The topography map (AFM) of the same area as that in (a-d & f-k). (f-k) The same sample as in (a-d) in a decreasing field after saturating in a 1.5 kOe field. (l-o) The Pt (3 nm)/NCO (15 nm) sample with a decreasing magnetic field after saturating at 1 kOe field. (q-v) The same sample as in (l-o) after demagnetization at high temperature. The squares in (n, t, & u) indicate pinning centers. (p) topography map of the same area as in (l-o & q-v).

**Figures 4(a-k)** show the "normal" processes of magnetization reversal in terms of magnetic domains in the sample Pt (4 nm)/NCO (50 nm) with thick NCO, measured by magnetic force microscopy (MFM). After saturation in -1.5 kOe field, nucleation of reversed domains occurs in the Pt (4 nm)/NCO (50 nm) sample, as indicated by the circle in **Figure 4 (a)**. As the field increases, the reversed domains expand via domain wall motion (**Figure 4 (b)**), before they coalesce and become a single domain in **Figure 4 (c-d)**. **Figures 4(f-k)** show similar processes with decreasing field after saturation in 1.5 kOe field, along with the AFM image of the same area in **Figure 4(e)**.

In contrast, these "normal" processes are absent in the sample of thinner NCO or Pt (3 nm)/NCO (15 nm). After saturation in 1 kOe field, MFM images were taken with an increasing field with a step of 10 Oe (see **Figure 4(l-o)**). There is no evidence of nucleation of reversed magnetic domains, except for a pinning site indicated by the square in **Figure 4(n)**. The missing domain structures were further confirmed by dedicated small area scans of 1 × 1 and 4 × 4 μm$^2$ (**Figure S7a** and **Figure S7b**, Supporting Information).

On the other hand, magnetic domains and domain-wall motion do appear in the same region (see topography map in **Figure 4(p)**) of the Pt (3 nm)/NCO (15 nm) sample after zero-field cooling, similar to that in the Pt (4 nm)/NCO (50 nm) sample shown in **Figure S7c**, Supporting Information. These domains gradually merge as the field is reduced in decreasing magnetic field, as shown in **Figure 4(q-v)**. A similar behavior is observed in another Pt (2 nm)/NCO (15 nm) in an increasing magnetic field (**Figure S7d**, Supporting Information), demonstrating the reproducibility of the results. Once the domain structure is wiped out by the magnetic field, subsequent MFM measurements for magnetization reversal show results similar to that in **Figure 4(l-o)** again, i.e., absent of domain structures. This behavior indicates that nucleation of reversed domains does occur during magnetization reversal in **Figure 4(l-o)**, but MFM (resolution ≈ 30 nm as limited by the tip size[19]) may not be able to detect the domain structures.

We further imaged the spin textures in Pt/NCO films using nitrogen-vacancy scanning probe microscope (NV-SPM)[20–22] (see **Figure S8** and **Figure S9** , Supporting Information). There is an apparent magnetic contrast (±120 mT) from the Pt (5 nm)/NCO (12 nm) sample in applied magnetic field in the range of 13–37 Oe, a possible indication of the presence of spin textures (see **Figure S8c** and **Figure S8e**). Similar results are obtained on Pt (5 nm)/NCO (14 nm) sample in a magnetic field of 50 Oe with a few larger features (**Figure S9a**) arising from topography (**Figure S9b**). In NV-SPM, the spatial resolution is given by the distance of the implanted NV defect to the sample surface (~50 nm in our setup).[20,21,23] Therefore, it is not possible to observe directly the topological spin textures like the nucleation centers in Pt/NCO with size < 50 nm using NV-SPM if the population density of nucleation centers is too high.

The fact that the magnetic domain structure is observed after zero-field cooling but not during nucleation process of the magnetization reversal in the Pt (3 nm)/NCO (15 nm) sample by scanning problem microscopy, including MFM and NV-SPM, is consistent with densely packed small nucleation centers that may be too small for the resolution.

To gain insight into the effect of DMI on the nucleation process in the Pt/NCO heterostructure, we carried out micromagnetic simulation on the magnetization reversal process. The topological charge (or skyrmion number) during the magnetization reversal and the corresponding magnetization with the DMI strength $D$ = 0.3 mJ m$^{-2}$ is shown in **Figure S10b**, Supporting Information as an example. The results from simulation agree with the experimental observation qualitatively. Starting from saturation in a positive field, nucleation occurs before the field reaches coercivity $H_C$, causing a dip in the topological charge, which is proportional to THE.

After the magnetization changes sign, the topological charge reduces quickly to zero. An important finding is that larger $D$ value leads to larger topological charge in the nucleation process (**Figure S10a**, Supporting Information), which confirms that DMI promotes nucleation and increases the number of nucleation centers. However, larger $D$ value makes it harder to saturate the magnetization and the nucleation process starts before zero field. Overall, the value $D$ = 0.1-0.5 mJm$^{-2}$ appears to match experimental observation most reasonably. We note that the $D \sim 0.1$ mJ m$^{-2}$ value is larger compared with other oxide interfaces[24–27] but smaller than that in heavy metal/ferromagnetic metal interfaces.[28,29] This may be related to the large orbital moment of Co, which may couple to Pt and generate a strong DMI.

Table 1. Normalized topological Hall resistivity comparison: Pt/NCO with other system.

| System | Temperature range (K) | $\rho_{THE}^{Max}(n\Omega cm)$ | $\frac{\rho_{THE}^{Max}}{\rho_{xx}}(10^{-3})$ | $n_s(\mu m^{-2})$ | Type |
|---|---|---|---|---|---|
| MnSi | 28-29[30] | 4.5[30] | 0.04[31] | | Bulk |
| Ta/MnGa (1 nm) | 5-300[32] | 3000[32] | 6.7[32] | 10$^4$ | Interface |
| Pt/TmIG (2 nm) | 275-370[33] | 5[33] | 0.6[33] | | Interface |
| SRO/SIO/STO | 10-70[34] | 110[34] | 0.0004[34] | | Interface |
| SRO/STO | 100-135[35] | 70[35] | 0.3[36] | | Interface |
| (Ca,Ce)MnO$_3$ | 0-100[18] | 120000[18] | 0.12[18] | | Bulk |
| Pt (3 nm)/NCO (15 nm) (this work) | 2-350 | 5000 (NCO layer) | 11.6 (NCO layer) | 10$^4$ | Interface |

Next, we discuss the magnitude of THE which is related to the size of the nucleation centers. Table I shows the comparison of THE between Pt/NCO and other previously reported systems. THE in Pt/NCO exists in the range of measurement temperature 2-350 K, which is shared with systems with DMI like Ta/MnGa.[32] For the magnitude of THE resistivity, Pt/NCO is lower than that of (Ca,Ce)MnO$_3$.[18] On the other hand, since THE corresponds to deflection of charge carriers in emergent field $\vec{B}_e$, the corresponding resistivity is determined by the product of skyrmion population density $n_s$ ($\propto \vec{B}_e$) and the Hall constant ($R_0$) of the ordinary Hall effect which is proportional to the longitudinal resistivity ($\rho_{xx}$). To make the comparison more reflective on the nucleation or skyrmion population density $n_s$, we calculate normalized topological Hall resistivity, *i.e.*, the ratio between THE resistivity and longitudinal resistivity, and display in Table I. $\frac{\rho_{THE}^{Max}}{\rho_{xx}}$ for NCO (see Supporting Information Section **S12**) demonstrates a value that surpasses all other values in Table I,[18,30–36] suggesting high nucleation density and small nucleation sizes.

Analysis of THE in terms of emergent magnetic field[4] allows for the determination of charge carrier type and estimation of the population density and sizes of the nucleation centers. Starting from the magnetization-up state, an external magnetic field may reverse magnetization to the down state locally and generate a nucleation center. Correspondingly, $\vec{B}_e$ points down, according to Eq. (1) in Supporting Information Section **S12**. As shown in **Figure 2(c)**, measured $R_{xy}$ is positive in this case, suggesting that the charge carrier type is n-type, consistent with the result from the Hall constant measurement in high field (see **Figure S12**, Supporting Information). We estimated an emergent magnetic field $|\langle\vec{B}_e\rangle|$ as high as 30 T and the nucleation density is $7 \times 10^3$ μm$^{-2}$ (see Supporting Information Section **S12**) for the Pt (3 nm)/NCO (15 nm) sample. This estimation implies that the diameter of the nucleation centers is about 10 nm when tightly packed. This high density and small sizes of the nucleation centers explain why MFM and NV-SPM cannot observe domain structures in the nucleation process of magnetization reversal, considering spatial resolution. Lorentz-transmission electron microscopy[37] may help in the observation of small (< 20 nm) skyrmions but its requirement of proper thin (transparent) sample makes it hard in our Pt/NCO/MAO system. The small topological spin textures in Pt/NCO are most likely related to its moderate exchange stiffness, strong PMA, the DMI introduced by the Pt/NCO interface, since these properties favor narrow domain walls and smaller nucleation centers.

The observation of large THE in the Pt (3 nm)/NCO (15 nm) sample is remarkable, because THE has been typically observed in systems with very thin (few nm) magnetic film.[38,39] Nevertheless, the interfacial nature of the THE in Pt/NCO is supported by the reduced signal in the Pt (4 nm) / NCO (50 nm) sample (**Figure S11**, Supporting Information) compared with that in Pt (3 nm) / NCO (15 nm), as well as the absence of THE in single-layer of NCO or Pt[15] and in Pt/Cu/NCO (**Figure S3**, Supporting Information). We attribute the observation of THE with NCO thicker than 10 nm to the large exchange length which determines the vertical spin correlation (see Supporting Information Section **S13**).

In conclusion, our investigation of Pt/NCO heterostructures reveals a giant THE across a wide temperature range. The presence of THE before $H_C$ is indicative of its relationship with magnetic nucleation. Micromagnetic simulations qualitatively confirm enhanced nucleation population density by DMI. The estimated high population density of nucleation suggests small size about 10 nm at room temperature. Our study unveils novel spin transport phenomena in Pt/NCO heterostructures, highlighting their potential for spintronics applications.

## Methods

**Sample preparation.** Epitaxial NiCo$_2$O$_4$ (001) films of about 8 - 50 nm thickness were grown on MgAl$_2$O$_4$ (001) substrates with a pulsed laser deposition system using a Nd-YAG solid state laser with 350 °C substrate temperature and 150 mTorr oxygen pressure. After the deposition of the NCO layer, a 3-4 nm Pt layer was deposited in-situ in $10^{-7}$ torr vacuum. Hall bar of size 200 $\mu m \times$ 160 $\mu m$ was made using photolithography and reactive ion etching.

**X-ray structural characterization**. Crystal structural and thickness of the films were measured using x-ray diffraction (XRD) and x-ray reflectivity (XRR) (**Figure S1**, Supporting Information) with a Rigaku SmartLab diffractometer ($\lambda = 1.54$ Å).

**Magneto-optical Kerr effect (MOKE) and magneto-transport measurements**. Out-of-plane magnetization was studied using a home-built MOKE system with photo elastic modulator (PEM) in a polar configuration in a Janis close-cycle optical cryostat system (20-320 K) and a GMW Dipole Magnet. The magneto-transport measurement was carried out in the same cryostat and magnet system as that for the MOKE measurements simultaneously. Additional magneto-transport measurements at temperature ranges of 2-20 K and 320-370 K were conducted using a Physical Properties Measurement System (PPMS) Dynacool system from Quantum Design.

**Magnetic force microscopy (MFM).** Magnetic domains structures of the films were measured using a Digital Instruments D3100 magnetic force microscopy. All the MFM images were obtained at the same scanning height of 50 nm.

**Magnetometry.** Magnetization of NCO films have been measured using superconducting quantum interference device (SQUID) from Quantum Design between 20 and 400 K. The *M-T* measurement was conducted during warming in a 5 kOe out-of-plane magnetic field. The *M-H* measurement was carried out at 20, 300, and 385 K.

**Micromagnetic simulation.** Micromagnetic simulation was carried out to study the magnetization reversal process with GPU-accelerated micromagnetic package, Mumax3, assuming a system size 1 μm × 1 μm × 14 nm. We adopt the experimental values for magnetization $M_s$ = 250 kAm$^{-1}$, $K_1$ = 0.21 MJ m$^{-3}$ at RT, and exchange stiffness $A_e$ = 4 pJ m$^{-1}$ estimated using mean-field theory based on the Curie temperature [see Supporting Information Section **S13**]. To account for the randomness of the Co and Ni valency and site occupation, we introduced exchanged coupled grains with sizes of about 10 nm with slightly varying anisotropy and coercivity.

## Supporting Information

The Supporting Information is available free of charge via the internet at http://pubs.acs.org.

Acknowledgements
This work is primarily supported by the NSF/EPSCoR RII Track-1: Emergent Quantum Materials and Technologies (EQUATE) Award OIA-2044049. A.L. acknowledges NSF award# 2328822 for support on magnetic imaging. Z.X. acknowledges funding from the European Union's Horizon 2020 research and innovation programme under the Marie Skłodowska-Curie grant agreement No 955671 on magnetic imaging. A.L. and X.X. acknowledges partial support from the UNL Grand Challenges catalyst award entitled "Quantum Approaches addressing Global Threats". The research was performed in part in the Nebraska Nanoscale Facility: National Nanotechnology Coordinated Infrastructure and the Nebraska Center for Materials and Nanoscience (and/or NERCF), which are supported by NSF under Award ECCS: 2025298, and the Nebraska Research Initiative.

## Supplementary Information

### S1. Thickness measurement

X-ray reflectivity (XRR) measurements were conducted to determine the thickness of Pt and NCO layers on the sample (see **Figure S1**). The XRR fitting analysis yielded the NCO and Pt thickness 15 and 3 nm with Pt and NCO roughness 0.2 and 0.5 nm respectively.

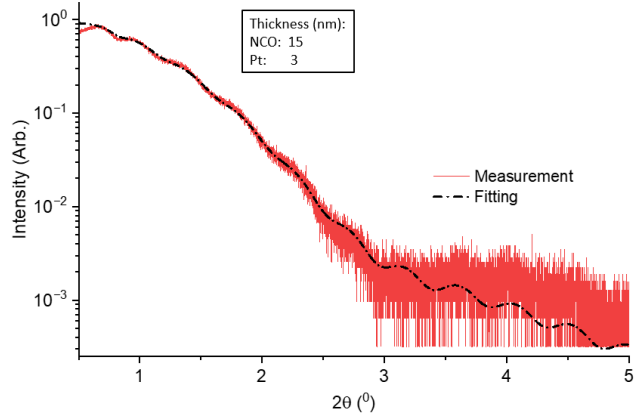

**Figure S1.** X-ray reflectivity scan (red) of the sample with fitted curve (black).

### S2. Hall measurement on NCO/MAO (001)

To investigate whether the NCO/MAO (001) without Pt layer exhibit THE like features or not, we conducted Hall measurements on various thickness as illustrated in **Figure S2**. Notably, our findings align with previous reports,[8,40] i.e., no THE-like characteristics such as dips or peaks are observed.

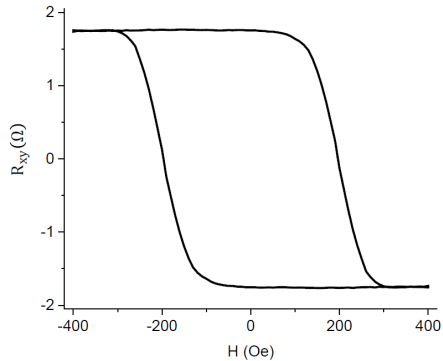

**Figure S2.** Hall measurement of NCO (12 nm)/MAO (001) using van der Pauw (vdP) method.

### S3. Hall measurement on Pt/Cu/NCO/MAO (001)

We prepared a sample with Cu at the interface between Pt and NCO and measured Hall resistance (see Fig. R4). We observed AHE loop but without obvious THE peaks near the coercive field. Note that this sample has a thinner NCO layer (≈10 nm) than the main sample discussed in main text, so the $T_c$ and coercive field are also slightly lower, consistent with previous work[8].

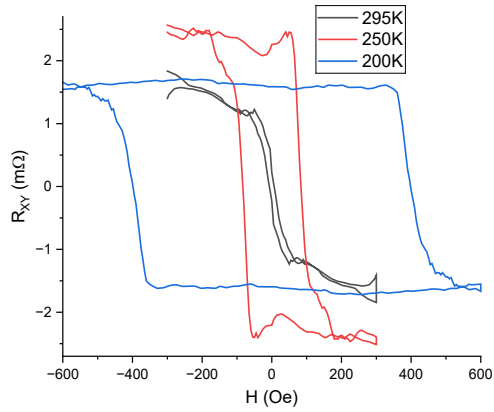

**Figure S3.** Hall measurement on Pt/Cu/NCO (10 nm)/MAO (001) at different temperatures.

## S4. Hall measurements at different temperatures

Temperature dependent Hall measurements were conducted using cryostat over the 20 - 300 K range and Quantum Design PPMS Dynacool system for low (2 – 20 K) and high (320 – 370 K) temperature ranges. Measurement exhibits similar behavior at all temperature range from 2-350 K, **Figure S4**. However, with temperatures above 300 K, a small peak appeared alongside the dip and increased with the rising temperature and disappearing close to the $T_C$.

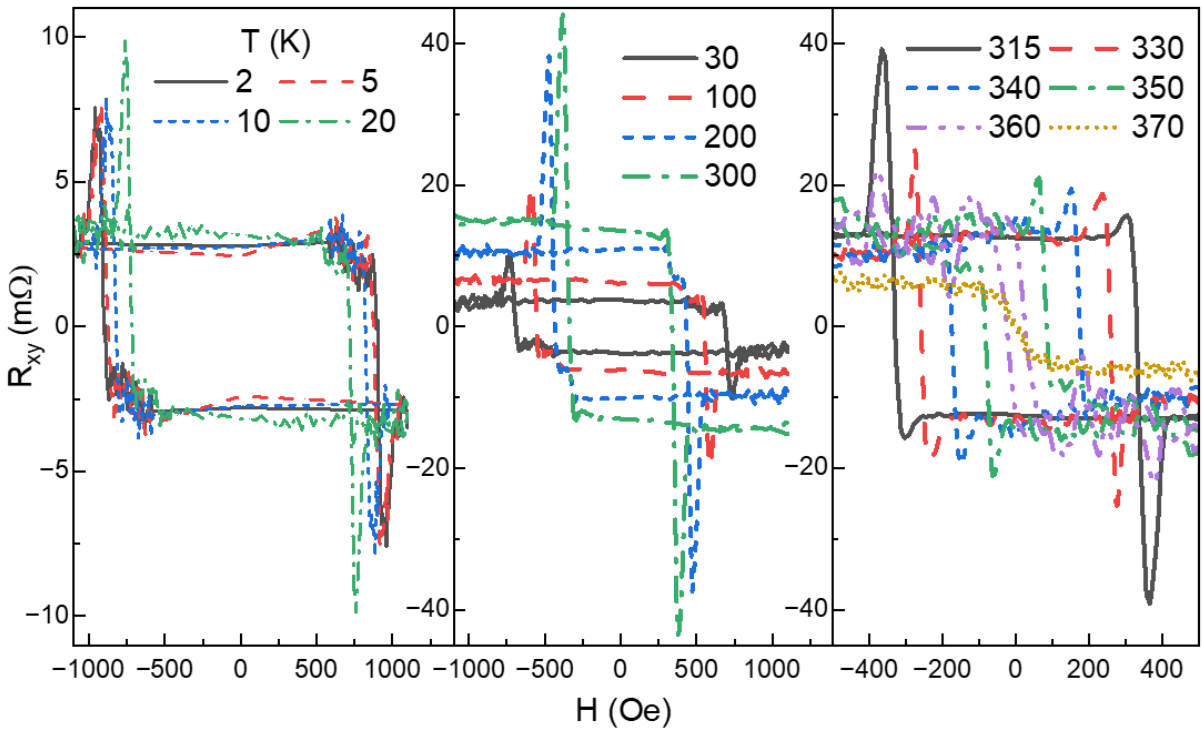

**Figure S4:** Hall measurements at different temperatures.

## S5. Multiple magnetic layers

The small steps observed in both Hall and magneto-optical Kerr effect (MOKE) measurements (see **Figure S5**) can be attributed to the sample preparation process. During the wet etching process used to create Hall bar structure, Pt is completely etched away, but only ~3 nm of NCO layer outside the Hall bar is removed, possibly contributing to these steps. This assumption is supported by considerations of $H_C$ dependence on thickness and temperature as reported in previous work,[8] and our own observation with samples of various thickness sample also reveal that the $H_C$ gets smaller for thinner sample. In addition, it is well-known that thinner NCO has lower $T_c$. It is clear from **Figure S5** that there is only one magnetic component at 300 K, suggesting that the residual NCO outside the Hall bar is not magnetic at 300 K. So, the step in MOKE comes from contribution from laser reflected from NCO outside of Hall bar. This impact is reflected in Hall measurement as well, given the semiconducting nature of NCO outside the Hall bar due to minor leaking current.

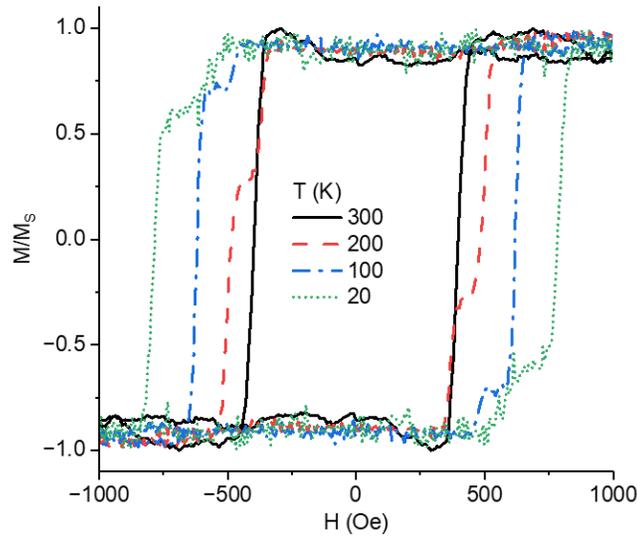

**Figure S5:** MOKE field dependence of normalized magnetization (*M-H*) loops at various temperatures.

Additionally, THE-like features in principle can be attributed to multiple AHE channels arise from distinct magnetic layers[17,41] as we possess two different magnetic layers with different magnetic properties. However, to have topological like peaks, two different magnetic layers must have positive and negative AHE with different magnitudes of $H_C$. However, the AHE sign for NCO remains the same near room temperature over a wide thickness range[8]. In fact, at 300 K, although THE is close to the maximum, there is only one magnetic component (see **Figure S5**). This nullifies the multichannel origin of THE.

## S6. Normalized topological Hall resistivity

In the main text, we presented a portion of $R_{xy}$ map, which includes AHE as well. The graph displayed here **Figure S6** represents the complete map of normalized topological Hall resistivity after extraction subtraction of AHE contributions for both increasing and decreasing magnetic field directions as indicated by black arrow along the *H*-axis. Note that the contribution to THE originates from NCO layer and normalized resistivity calculated is only for NCO layer which is $\frac{\rho_{xy,NCO}}{\rho_{xx,NCO}} = \beta \frac{\rho_{xy}}{\rho_{xx}}$, where $\beta \equiv 1 + \frac{d_{Pt}}{d_{NCO}} \frac{\rho_{NCO}}{\rho_{Pt}}$ (see Supplementary Information **Section S12** for detail).

Temperature dependent resistivity of 16 nm NCO[8] and 3.8 nm Pt[15] is used to calculate $\beta$ and hence $\frac{\rho_{xy,NCO}}{\rho_{xx,NCO}}$. This map effectively illustrates the region where the THE is observed, and notably, THE exhibits asymmetry across the entire temperature range.

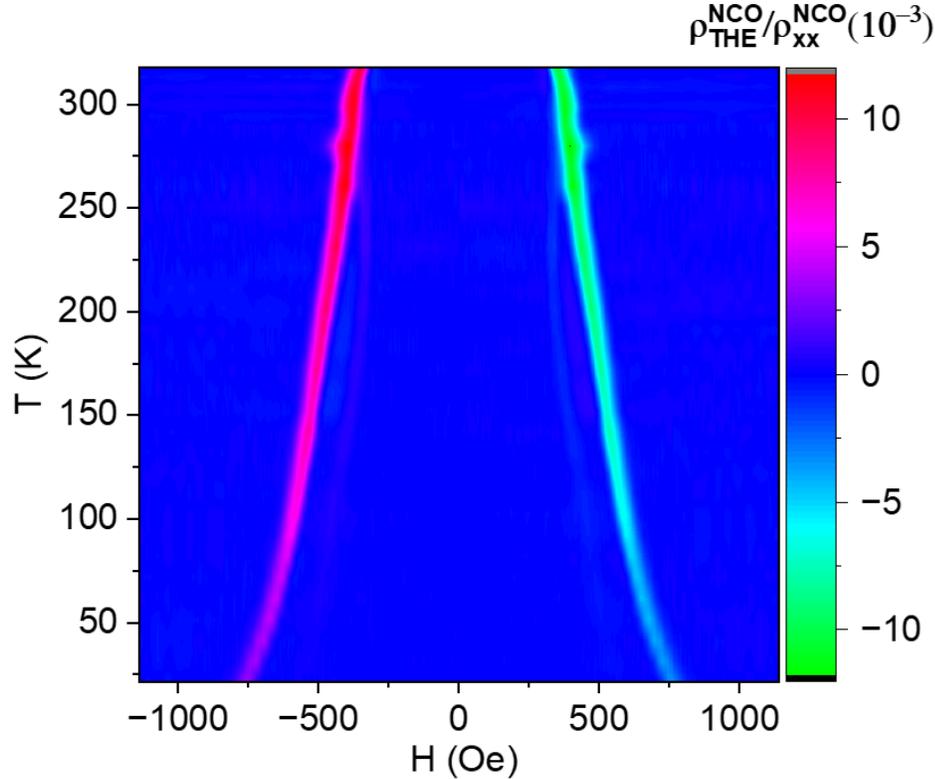

**Figure S6:** Contour map of normalized topological Hall resistivity in the plane (with respect to longitudinal resistivity) of magnetic field and temperature. In this single graph Data from 0 to 1100 Oe are consolidated, covering increasing and decreasing field directions starting from magnetization saturation point. Red and green color shows THE while when the field is decreasing and increasing respectively. The blue area shows the region with no THE.

**S7. Magnetic force microscopy (MFM) measurements**
After each MFM scan of area 20× 20 μm² following saturation, as shown in **Figure 4 (l-o)**, additional scans of area 4 × 4 μm² and 1 × 1 μm² were conducted with decreasing magnetic field steps of 10 Oe. No observable domains are observed across all scans. The representative images of topography map (AFM) and MFM at -243 and -283 Oe are shown in **Figure S7 (a and b)**. Note that the $H_C$ (≈ 250 Oe) and minimum saturating magnetic field (≈ 310 Oe) from MFM is smaller than that for Hall or MOKE measurements at room temperature. This discrepancy might be because of the slightly different magnetic field calibration in MFM. **Figure 7 (c)** illustrates the evolution of magnetic domains in zero field cooled Pt (4 nm) / NCO (50 nm) under a decreasing magnetic field. The thicker NCO (50 nm) layer exhibit significantly smaller magnetic domains compared to the 15 nm NCO. **Figure 7 (d)** presents the evolution of magnetic domains in another zero-field cooled sample, Pt (2 nm) / NCO (15 nm) under a increasing magnetic field, confirming that magnetic domain size (~2 μm) and behavior consistent with the main sample discussed in the main text.

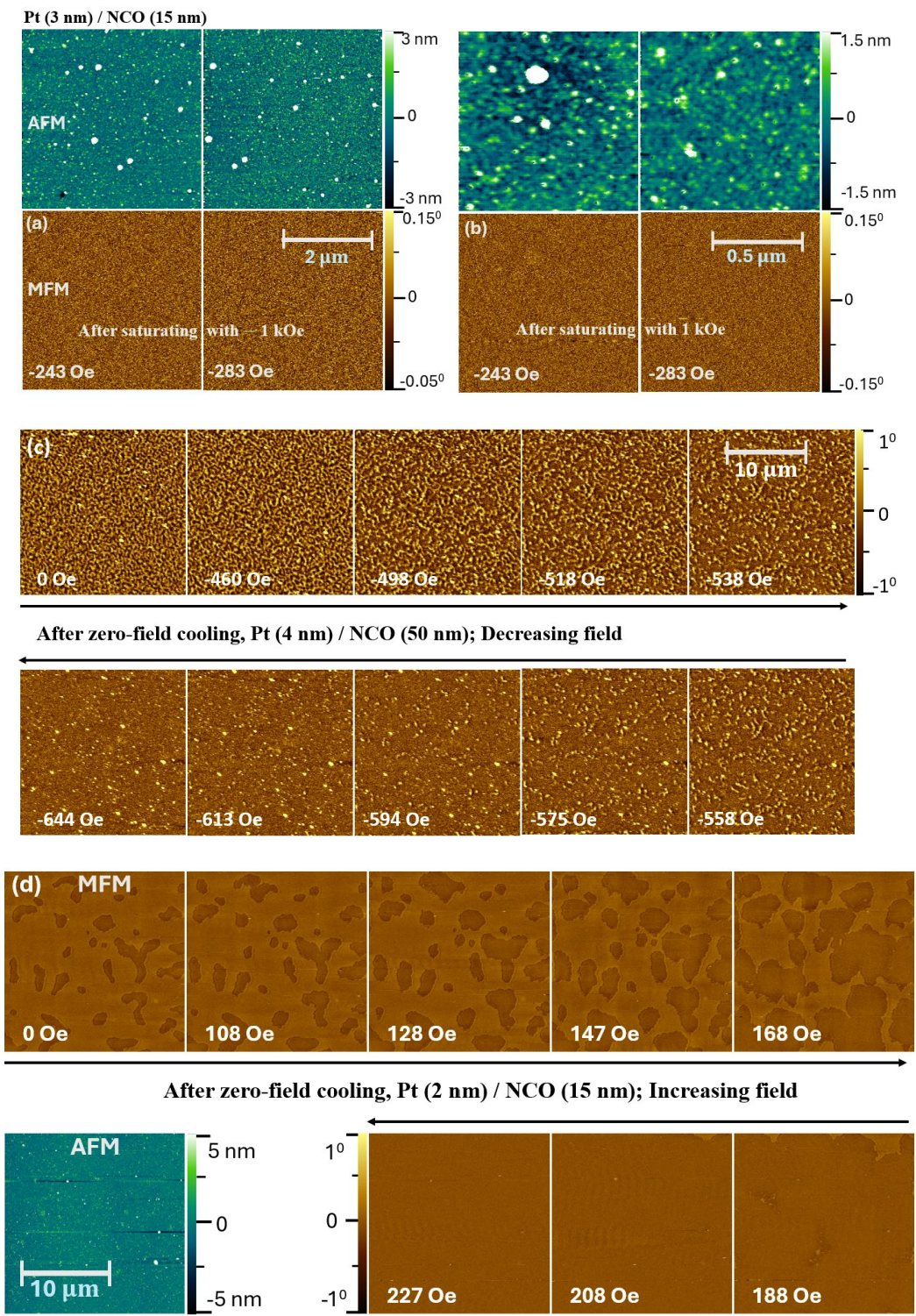

**Figure S7**: Representative AFM (top) and MFM (bottom) scans of area (a) 4 × 4 μm² and (b) 1 × 1 μm² at -243 and -283 Oe field after magnetic saturation in 1.5 kOe. (c) shows MFM images (20 × 20 μm²) of zero field cooled Pt (4 nm) / NCO (50 nm) with a decreasing magnetic field. (d) shows the MFM images (20 × 20 μm²) of zero field cooled Pt (2 nm) / NCO (15 nm) with an increasing magnetic field, along with the corresponding AFM images (bottom left).

## S8. Nitrogen-Vacancy scanning probe microscopy of NCO/Pt films

Apparent contrasts in MFM due to magnetic forces are often contaminated by other long-range forces associated, for instance, with surface charges that make it hard to quantitatively interpret the measured magnetic signals.[42] Recently, a new technique has emerged for measuring magnetic fields at the nanometer scale based on optical detection of the electron spin resonances of nitrogen vacancy (NV) centers in diamond.[43] Negatively charged NV centers are spin 1 defects with a spin-triplet ground state, featuring a zero-field splitting, $D$ = 2.87 GHz, between states $m_s$ = 0 and $m_s$ = ±1. Laser illumination (532 nm) produces a spin-conserving transition to the first excited triplet state, which in turn leads to fluorescence (650-750 nm), **Figure S8a**. Intersystem crossing to metastable singlet states takes place preferentially for NV centers in the $m_s$ = ±1 states allowing optical readout of the spin state via optically detected magnetic resonance (ODMR).[43] The application of a magnetic field breaks the degeneracy of the $m_s$ = ±1 state and leads to a pair of transitions whose frequencies depend on the magnetic field component along the NV symmetry axis, **Figure S8b**. NV magnetometry is now widely used to detect static and dynamic magnetic stray fields from solid-state systems in a wide range of temperature (0.3-600 K), opening up new frontiers in condensed matter research.[44]

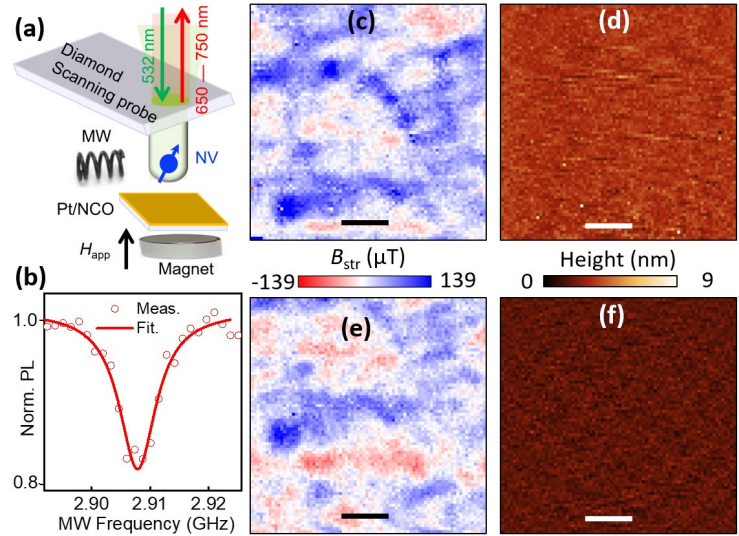

**Figure S8.** Nitrogen-vacancy scanning probe microscopy (NV-SPM) measurements Pt/NCO (12 nm). **(a)** Schematic of NV-SPM. (b) Two-dimensional NV magnetic stray field ($B_{str.}$) image of a selected NCO/Pt region at $H_{app}$ of 13 Oe (c) and 37 Oe (e). The NV standoff is ~ 50 nm. AFM image of the same region in (c) and (e) at $H_{app}$ of 13 Oe (d) and 37 Oe (f), respectively. The scale bar in **c-f** is 200 nm.

To map spin textures in Pt/NCO films, we used a custom NV-scanning probe microscope (NV-SPM), see its schematic in **Figure S8a**. More specifics about the NV-SPM microscope can be found in reference [21]. ODMR imaging is performed to measure the stray field, $B_{str}$, generated by the spin textures at a certain distance, $d_{NV}$, above the Pt/NCO film. **Figures S8c and S8e** show $B_{str}$ maps of a selected Pt/NCO (12 nm) region at an applied magnetic field $H_{app}$ of 13 Oe and 37 Oe, respectively, and at standoff of ~ 50 nm which is inherent to the non-contact shear mode atomic force feedback loop. Note that there is a slight offset in the scanned area due to a small sample drift. In each pixel of the NV magnetic images, we measured NV ODMR spectrum (**Figure S8b**) to extract $B_{str}$ produced by spin textures. There is an apparent magnetic contrast (±120 µT) from the Pt/NCO film that does not change significantly with $H_{app}$ in the range of 13-37 Oe. The magnetic contrast of possible spin textures in Pt/NCO does not correlate with the simultaneously acquired topography in **Figure S8d** and **Figure S8f**. Similar results are obtained on Pt/NCO (14 nm) film at $H_{app}$ of 50 Oe with a few big features (**Figure S9a**) arising from topography (**Figure S9b**). Since our NV-SPM has a limit of the maximum applied magnetic field ~ 200 Oe, we performed MFM measurements on the same film around the THE maximum peak and see no apparent phase contrast (**Figure S9c**) with a few features coming from the topography (**Figure S9d**).

In NV-SPM, the spatial resolution is given by the distance of the implanted NV defect to the sample surface (~50 nm in our setup).[23] Therefore, it is not possible to confirm the presence of topological spin textures such as skyrmions in Pt/NCO films with size ~ 10 nm, deduced from THE measurements. Though, the high magnetic stray field contrast in the NV images (e.g., in **Figures S9c** and **S9d**) suggests the presence of spin textures in Pt/NCO films.

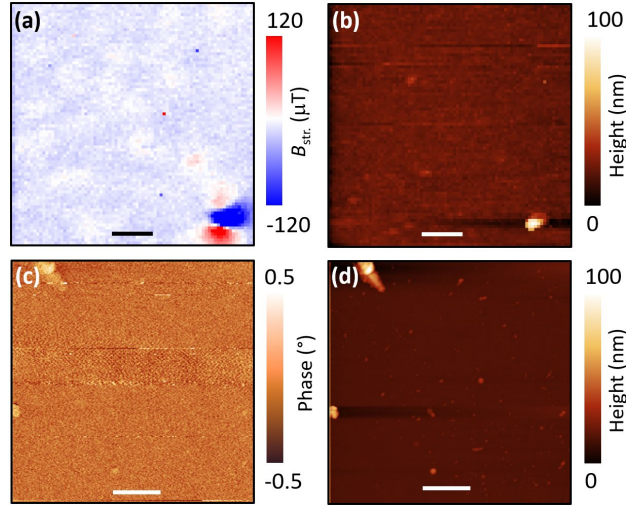

**Figure S9. (a)** 2D NV magnetic stray field ($B_{str.}$) image of a selected Pt/NCO (14 nm) region at $H_{app}$ of 50 Oe. The NV standoff is ~ 150 nm. **(b)** The corresponding AFM image of the region in (a). The scale bar in **a-b** is 200 nm. MFM **(c)** and AFM **(d)** of another Pt/NCO (14 nm) region performed at $H_{app}$ of 290 Oe. The scale bar in **c-d** is 1 mm.

## S9. Micromagnetic simulations
For micromagnetic simulations we considered the energy function with DMI interaction. The local configuration $\boldsymbol{M}(\boldsymbol{r}) = M_s \boldsymbol{m}(\boldsymbol{r})$ with $\|\boldsymbol{m}\| = 1$ is determined using energy function given by:[45,46]

$$\mathcal{E} = \int \left\{ A_e \left[ \nabla \left( \frac{\boldsymbol{M}}{M_s} \right) \right]^2 + D \left[ m_z (\nabla \cdot \boldsymbol{m}) - (\boldsymbol{m} \cdot \nabla) m_z \right] - K_1 \frac{(\boldsymbol{n}.\boldsymbol{M})^2}{M_s^2} - \mu_o \boldsymbol{M}.\boldsymbol{H} - \frac{\mu_o}{2} \boldsymbol{M}.\boldsymbol{H}_d(\boldsymbol{M}) \right\} dV.$$

The 1$^{st}$ term is the exchange interaction with exchange stiffness $A_e$ parameterizes the interatomic exchange energy $\mathcal{E}_A = A_e (\nabla \boldsymbol{S})^2$, the 2$^{nd}$ term is interfacial DMI interaction arises due to spin orbit interaction due to the presence of Pt film right next to NCO film,[47] $K$ is the uniaxial perpendicular anisotropy assumed to be along the $c$-axis in the z-direction. There are two magnetostatic terms in Eq. (1), namely the Zeeman interactions with the external magnetic field $H$, and the magnetostatic self-interaction energy described by the demagnetizing field $\boldsymbol{H}_d(\boldsymbol{M})$. To explore the topological spin textures and THE numerically, the topological charge ($Q$) was extracted from the spin structure to quantify the spin textures.

We employed a GPU-accelerated micromagnetics package, Mumax3, to study the hysteresis loop and topological charge of the NCO thin film with DMI interaction under the applied magnetic field. The NCO film has regions in nm with different Ni and Co concentration. These grains represent small magnetic regions in the film with slightly different materials parameters. For this purpose, to match real samples (assuming your real sample is polycrystalline) we used ext_makegrains function based on Voronoi tessellation to add features like grains and then vary the material parameters slightly by grains.[48] This helps smooth things out, as some grains have slightly lower (or higher) anisotropy/coercivity.

The exchanged coupled grains have sizes of about 10 nm, and the total size of the simulated system, 1 μm × 1 μm × 14 nm. We have used a computational cell size of 3 nm, which is well below the exchange length $l_{ex}$. For the study of spin textures and magnetization reversal in NCO/Pt film temperature-dependent micromagnetic effects are included in the lowest order that is, by considering the intrinsic materials parameters $M_s$, $K_1$, and $A$ as temperature dependent. This approach accounts for the atomic spin disorder. In our simulations, we have taken values of $M_s$ = 250 kAm$^{-1}$, $K_1$ = 0.21 MJm$^{-3}$, and $A$ = 4 pJm$^{-1}$ (see Section S12) and vary DMI interaction $D$ from 0.1 to 0.5 mJm$^{-2}$. A larger $D$ value correlates with large Skyrmion numbers or topological charge during the nucleation process. This correlation extends for Skyrmion numbers after $H_C$ as well and becomes significant at $D$ = 0.5 mJ m$^{-2}$ (see **Figure S10**).

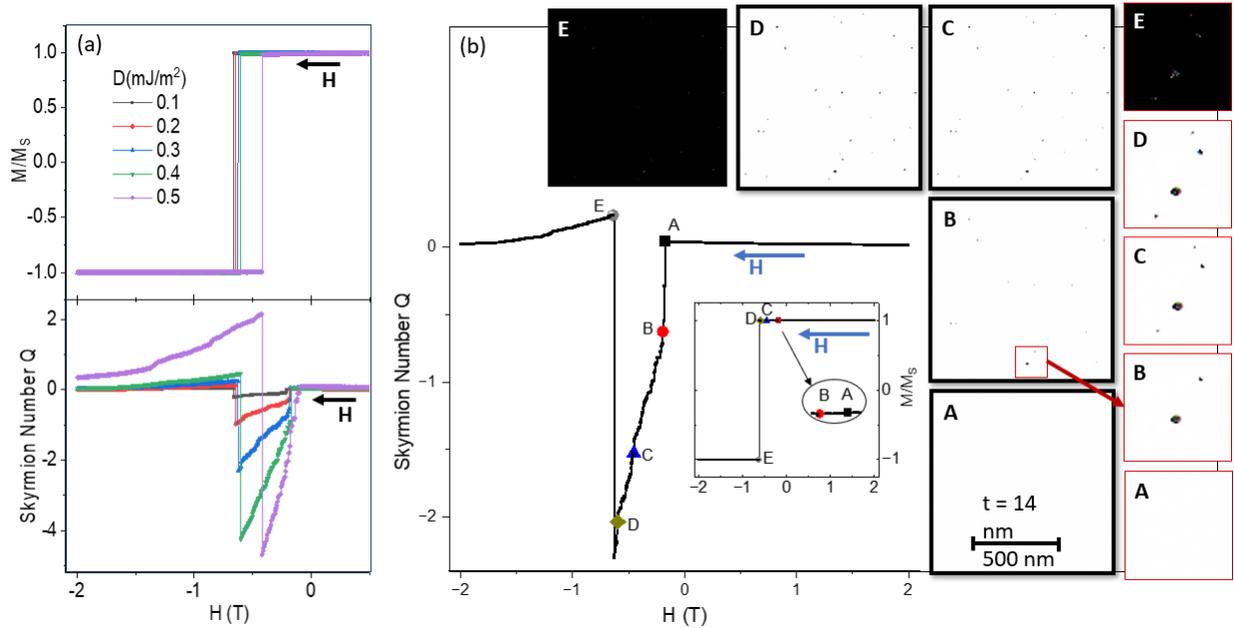

**Figure S10. (a)** Predicted normalized magnetization (top) and Skyrmion number or topological charge (bottom) as the magnetic field decreases from 2 T at various DMI constants using numerical simulation for 14 nm NCO thin film. **(b)** Numerical simulation of Skyrmion numbers during the magnetization reversal of a 14 nm NCO thin film with DMI introduced by Pt/NCO interface ($D$ = 0.3 mJ m$^{-2}$), while the inset shows the *M-H* loop. The images in black depict the predicted domain structure of 1 × 1 μm$^2$ area at various fields with decreasing magnetic field (*H*). Additionally, images in red square are a zoomed view of specific regions within the black square images.

**S10. Hall measurement of thick NCO**

Similar Hall and MOKE measurements were conducted for a Pt (4 nm) / NCO (50 nm) sample with an NCO thickness of 50 nm. This system showed a similar behavior, but the topological Hall resistance is approximately 10 times weaker than that of the Pt (3nm) / NCO (reported 15 nm) NCO sample, see **Figure 2**. This observation further strengthens the hypothesis of an interfacial layer housing the Skyrmions.

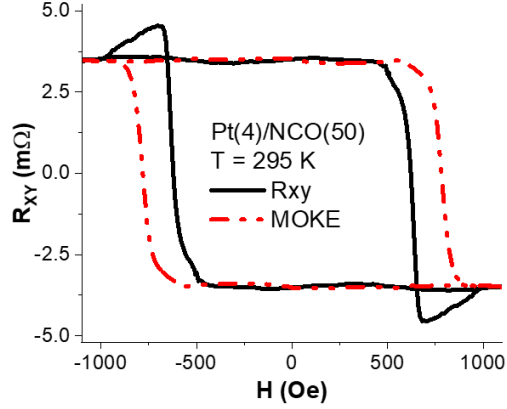

**Figure S11:** Measured Hall (vdP method) and MOKE curves of a Pt (4 nm) / NCO (50 nm) / MAO (001) film at 295K.

### S11. Hall constant measurement

This measurement serves the purpose of calculating the Hall constant ($R_0$) of our system. In high field region, $R_{xy}$ exhibits linear relationship with applied magnetic field (**Figure S12**). The Hall constant ($R_0$) is determined from the slope of this linear relationship and can be expressed as the following (see Supplementary Information **Section S12**).

$$R_0^{NCO} = \left(1 + \frac{\rho_{NCO}}{\rho_{Pt}} \frac{d_{Pt}}{d_{NCO}}\right)^2 d_{NCO} \frac{dR_{xy}}{dB_z} = \eta \left(1 + \frac{\rho_{NCO}}{\rho_{Pt}} \frac{d_{Pt}}{d_{NCO}}\right)^2 d_{NCO}$$

where $d_{NCO}$ and $d_{Pt}$ represent are the thickness of the NCO and Pt layers respectively, $\rho_{NCO}$ and $\rho_{Pt}$ are the resistivity of NCO and Pt respectively and $\eta \equiv \frac{dR_{xy}}{dB_z}$ is the in slope in the high field region.

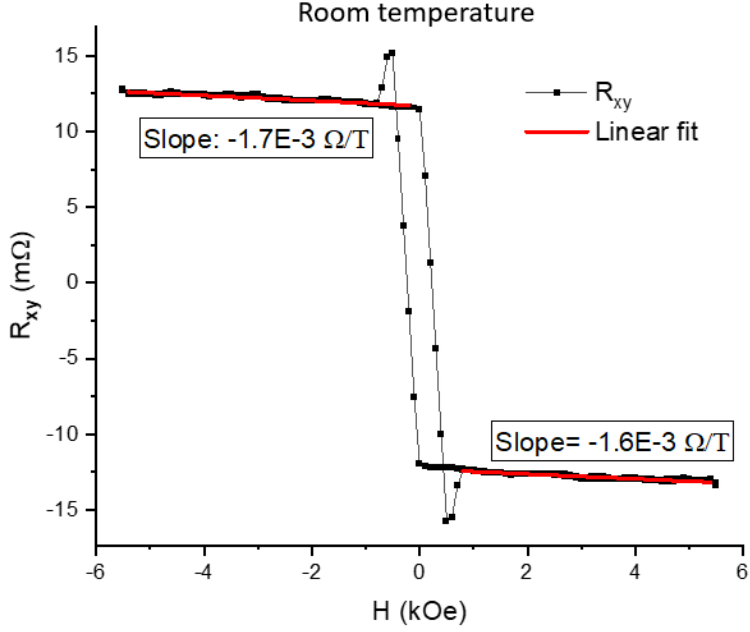

**Figure S12** Hall measurement of the sample at high magnetic field. Red line shows the fitted line for $R_{xy}$ vs H graph at high field region.

### S12. Skyrmion density derivation and estimation by emergent-field model
- **Longitudinal transport**

Given the bilayer system with both layers being conducting, the current through the sample can be written as:
$$I = j_x w(d_{NCO} + d_{Pt}),$$
where $w$ is the width (transverse direction) of the Hall bar, $d_{Pt}$ and $d_{NCO}$ are the thickness of the Pt layer and the NCO layer respectively.

For longitudinal transport, we have resistance
$$R_{Pt,L} = \rho_{Pt} \frac{L}{w d_{Pt}}, R_{NCO,L} = \rho_{NCO} \frac{L}{w d_{NCO}},$$
where $L$ is the length (longitudinal direction) of the Hall bar.

For the current through NCO layer:
$$I_{x,NCO} = I_x \frac{R_{Pt,L}}{R_{Pt,L} + R_{NCO,L}} = I_x \frac{\rho_{Pt} \frac{L}{w d_{Pt}}}{\rho_{Pt} \frac{L}{w d_{Pt}} + \rho_{NCO} \frac{L}{w d_{NCO}}} = \frac{I_x}{\beta}$$

$$j_{x,NCO} = \frac{I_{x,NCO}}{w d_{NCO}} = \frac{\left(1 + \frac{d_{Pt}}{d_{NCO}}\right)}{\beta} j_x$$

where $\beta \equiv 1 + \frac{d_{Pt}}{d_{NCO}} \frac{\rho_{NCO}}{\rho_{Pt}}$, $j_x$ is the longitudinal current density, $\rho$ are the longitudinal resistivity of corresponding labeled layers.

- **Transverse transport**

The Hall effect in NCO generates electro motive force (EMF) $V_{y,NCO}$. The introduction of Pt layer may generate a circuit in which NCO acts as the voltage source and drives the current in Pt and

NCO. The actual measured voltage is then reduced due to the "internal" resistance of NCO. We therefore have:

$$V_y = V_{y,NCO} \frac{R_{Pt,T}}{R_{Pt,T} + R_{NCO,T}} = V_{y,NCO} \frac{\rho_{Pt} \frac{w}{L d_{Pt}}}{\rho_{Pt} \frac{w}{L d_{Pt}} + \rho_{NCO} \frac{w}{L d_{NCO}}} = \frac{V_{y,NCO}}{\beta}$$

$$V_{y,NCO} = \beta V_y$$

The transverse resistivity is calculated as:

$$\rho_{xy,NCO} = \frac{E_{y,NCO}}{j_{x,NCO}} = \frac{V_{y,NCO}}{w j_{x,NCO}}.$$

Plug in $j_{x,NCO} = \beta \left(1 + \frac{d_{Pt}}{d_{NCO}}\right) j_x$ and $V_{y,NCO} = \frac{V_y}{\beta}$, we have

$$\rho_{xy,NCO} = \frac{V_{y,NCO}}{w j_{x,NCO}} = \frac{V_y \beta}{w \frac{\left(1 + \frac{d_{Pt}}{d_{NCO}}\right)}{\beta} j_x} = \frac{V_y}{w j_x} \frac{\beta^2}{\left(1 + \frac{d_{Pt}}{d_{NCO}}\right)} = \rho_{xy} \frac{\beta^2}{\left(1 + \frac{d_{Pt}}{d_{NCO}}\right)}$$

$$R_{xy,NCO} = \frac{V_{y,NCO}}{I_{x,NCO}} = \frac{\beta V_y}{\frac{I_x}{\beta}} = \beta^2 \frac{V_y}{I_x} = \beta^2 R_{xy}.$$

Regarding the normalized resistivity ratio:

$$\frac{\rho_{xy,NCO}}{\rho_{xx,NCO}} = \frac{\frac{E_{y,NCO}}{j_{x,NCO}}}{\frac{E_x}{j_{x,NCO}}} = \beta \frac{\rho_{xy}}{\rho_{xx}}$$

- ***Skyrmion density***

For the derivation of skyrmion density, we derive the ordinary Hall constant of NCO in terms of the slope of the field dependence of Hall resistance in the high field region. The ordinary Hall effect is described by the Hall coefficient:

$$R_0^{NCO} = \frac{1}{j_{x,NCO}} \frac{dE_{y,NCO}}{dB_z} = \beta^2 d_{NCO} \eta$$

where $\eta \equiv \frac{dR_{xy}}{dB_z}$ is the slope of $R_{xy}$ vs $B_z$ graph in high field region.

Since the topological Hall effect is the ordinary Hall effect because of emerging field ($B_e$), we can relate them:

$$R_0^{NCO} B_e = \rho_{THE}^{NCO}$$

Here we used polarization $P = 1$, since NCO is semi metallic[49].

$$\rightarrow B_e = \frac{\rho_{THE}^{NCO}}{R_0^{NCO}} = \frac{\frac{R_{THE}^{NCO}}{w} d_{NCO} L}{\beta^2 d_{NCO} \eta} = \frac{\frac{\beta^2 R_{THE}}{w} L}{\eta \beta^2} = \frac{R_{THE} L}{\eta w}.$$

The Skyrmions density is

$$n_s = \frac{e B_e}{h} = \frac{e}{h} \frac{R_{THE} L}{\eta w}.$$

Summarizing, we have the following useful relations

$$I_{x,NCO} = \frac{I_x}{\beta}$$

$$j_{x,NCO} = \frac{\left(1 + \frac{d_{Pt}}{d_{NCO}}\right)}{\beta} j_x$$

$$V_{y,NCO} = \beta V_y$$

$$\rho_{xy,NCO} = \rho_{xy} \frac{\beta^2}{\left(1 + \frac{d_{Pt}}{d_{NCO}}\right)}$$

$$R_{xy,NCO} = \beta^2 R_{xy}$$

$$\frac{\rho_{xy,NCO}}{\rho_{xx,NCO}} = \beta \frac{\rho_{xy}}{\rho_{xx}}$$

$$B_e = \frac{R_{THE} L}{\eta w}$$

$$n_s = \frac{e}{h} \frac{R_{THE} L}{\eta w}$$

When itinerant electrons traverse through the spin texture, under adiabatic conditions, their spins follow the direction of the local spins, i.e., the local spin eigenstates. For spin texture in the x-y plane, the itinerant electronic states acquire a Berry curvature perpendicular to the plane, as described by:[50]

$$\vec{\Omega} = \frac{1}{2} \hat{z}\, \hat{m} \cdot (\partial_x \hat{m} \times \partial_y \hat{m}) \qquad \text{Equation (1)}$$

where $\hat{m}$ is the local magnetization direction. The way the electronic spins follow the local spins can be treated using an emergent magnetic field which is proportional to the Berry curvature: $\vec{B}_e = \frac{\hbar}{e} \vec{\Omega}$. For non-coplanar spin textures such as skyrmions, the Berry curvature and emergent magnetic field (along the out-of-plane direction) are non-zero, which deflects the electrons and generates the topological Hall effect.

Skyrmion population density can be estimated from the magnitude of THE. Since each Skyrmions carries one magnetic flux quantum, average magnetic field (magnetic flux density) can be calculated as:

$$\langle \vec{B}_e \rangle = n_s \Phi_0 \qquad \text{Equation (2)}$$

where $\Phi_0$ is the magnetic flux of a Skyrmion whose magnitude equals magnetic flux quantum $\frac{h}{e}$, $h$ and $e$ are the Plank constant and the electronic charge. Hence, one has $|\langle \vec{B}_e \rangle| = n_s \frac{h}{e}$. Emergent magnetic field can be extracted from experimental measurements using (see Supplementary Information Section **S8**)

$$|\langle \vec{B}_e \rangle| = \frac{R_{THE} L}{\eta w} \qquad \text{Equation (3)}$$

where $R_{THE}$ is the THE resistance, $\eta$ is the slope of field dependence of $R_{xy}$ in the high field region (ordinary Hall effect, see **Figure S10**), $L = 200$ μm and $w = 160$ μm are length and width of the Hall bar skyrmion population density can then be estimated as:

$$n_s = \frac{e}{h} \frac{R_{THE} L}{\eta w},$$

Using the measured $\eta = 2 \times 10^{-3}$ $\Omega T^{-1}$ and $R_{THE} = 50$ $m\Omega$ at RT, the estimated $|\langle \vec{B}_e \rangle|$ is then 30 T and the population density is $7 \times 10^3$ μm$^{-2}$ for the Pt (3 nm)/NCO (15 nm) sample.

## S13. Estimation of magnetic parameters of NiCo$_2$O$_4$

- *Exchange stiffness*

In NiCo$_2$O$_4$, there are 8 magnetic Co and 8 magnetic Ni, on T and O sites respectively, where the T-O exchange (AFM) interactions dominate. The coordination number for both magnetic Co and Ni is 6. We then use the mean field theory to estimate the exchange interaction:

$$J = \frac{3k_B T_c}{2ZS(S+1)}$$

where $Z = 6$. For spin quantum number, we consider $S = 3.5/2$ for Co$^{2.5+}$ (3.5 µB), and $S = 1.5/2$ for Ni$^{2.5+}$ (1.5 µB). On average, $S = 1.25$. $T_c = 420$ K. Hence, we find $J = 5.15 \times 10^{-22}$ J.

Each magnetic ion occupies $v = \frac{(8.1 \times 10^{-10})^3}{16} = 3.32 \times 10^{-29}$ m$^3$. The effective lattice constant is then $a = v^{1/3} = 3.21 \times 10^{-10}$ m. The linear density of exchange (exchange stiffness) is then

$A_e = \frac{JS(S+1)}{a} = 4.5 \times 10^{-12}$ J m$^{-1}$ = 4.5 pJ m$^{-1}$.

- *Exchange length*

Exchange length is defined as $L_{ex} = \sqrt{\frac{A_e}{U_M}}$, where $A_e$ is the exchange stiffness, $U_M$ is the magnetic static energy. We can calculate the magnetostatic energy density using $U_M = \frac{\mu_o}{2} M_S^2$, where $M_S$ is the saturation magnetization. Using $M_S = 2.1$ $\mu_B$(f.u.)$^{-1}$, we find $U_M = 5.4 \times 10^4$ J m$^{-3}$ and $L_{ex} = 9.1$ nm.

**References**

(1) Nagaosa, N.; Tokura, Y. Topological Properties and Dynamics of Magnetic Skyrmions. *Nat Nanotechnol* **2013**, *8* (12), 899–911. https://doi.org/10.1038/nnano.2013.243.

(2) Braun, H. B. Topological Effects in Nanomagnetism: From Superparamagnetism to Chiral Quantum Solitons. *Adv Phys* **2012**, *61* (1), 1–116. https://doi.org/10.1080/00018732.2012.663070.

(3) Morshed, M. G.; Vakili, H.; Ghosh, A. W. Positional Stability of Skyrmions in a Racetrack Memory with Notched Geometry. *Phys Rev Appl* **2022**, *17* (6), 064019. https://doi.org/10.1103/PhysRevApplied.17.064019.

(4) Everschor-Sitte, K.; Sitte, M. Real-Space Berry Phases: Skyrmion Soccer (Invited). *J Appl Phys* **2014**, *115* (17), 172602. https://doi.org/10.1063/1.4870695.

(5) Ye, J.; Kim, Y. B.; Millis, A. J.; Shraiman, B. I.; Majumdar, P.; Teanovic, Z. Berry Phase Theory of the Anomalous Hall Effect: Application to Colossal Magnetoresistance Manganites. *Phys Rev Lett* **1999**, *83* (18), 3737–3740. https://doi.org/10.1103/PhysRevLett.83.3737.

(6) Bruno, P.; Dugaev, V. K.; Taillefumier, M. Topological Hall Effect and Berry Phase in Magnetic Nanostructures. *Phys Rev Lett* **2004**, *93* (9), 096806. https://doi.org/10.1103/PhysRevLett.93.096806.

(7) Xu, X.; Mellinger, C.; Cheng, Z. G.; Chen, X.; Hong, X. Epitaxial $NiCo_2O_4$ Film as an Emergent Spintronic Material: Magnetism and Transport Properties. *J Appl Phys* **2022**, *132* (2), 020901. https://doi.org/10.1063/5.0095326.

(8) Chen, X.; Zhang, X.; Han, M.; Zhang, L.; Zhu, Y.; Xu, X.; Hong, X. Magnetotransport Anomaly in Room-Temperature Ferrimagnetic $NiCo_2O_4$ Thin Films. *Advanced Materials* **2019**, *31* (4), 1805260. https://doi.org/10.1002/adma.201805260.

(9) Mellinger, C.; Waybright, J.; Zhang, X.; Schmidt, C.; Xu, X. Perpendicular Magnetic Anisotropy in Conducting $NiCo_2O_4$ Films from Spin-Lattice Coupling. *Phys Rev B* **2020**, *101* (1), 14413. https://doi.org/10.1103/PhysRevB.101.014413.

(10) Yoshimura, Y.; Kim, K. J.; Taniguchi, T.; Tono, T.; Ueda, K.; Hiramatsu, R.; Moriyama, T.; Yamada, K.; Nakatani, Y.; Ono, T. Soliton-like Magnetic Domain Wall Motion Induced by the Interfacial Dzyaloshinskii-Moriya Interaction. *Nat Phys* **2016**, *12* (2), 157–161. https://doi.org/10.1038/nphys3535.

(11) Pizzini, S.; Vogel, J.; Rohart, S.; Buda-Prejbeanu, L. D.; Jué, E.; Boulle, O.; Miron, I. M.; Safeer, C. K.; Auffret, S.; Gaudin, G.; Thiaville, A. Chirality-Induced Asymmetric Magnetic Nucleation in Pt/Co/AlOx Ultrathin Microstructures. *Phys Rev Lett* **2014**, *113* (4), 047203. https://doi.org/10.1103/PhysRevLett.113.047203.

(12) Benitez, M. J.; Hrabec, A.; Mihai, A. P.; Moore, T. A.; Burnell, G.; McGrouther, D.; Marrows, C. H.; McVitie, S. Magnetic Microscopy and Topological Stability of Homochiral Néel Domain Walls in a Pt/Co/AlOx Trilayer. *Nat Commun* **2015**, *6* (1), 8957. https://doi.org/10.1038/ncomms9957.

(13) Fert, A.; Cros, V.; Sampaio, J. Skyrmions on the Track. *Nat Nanotechnol* **2013**, *8* (3), 152–156. https://doi.org/10.1038/nnano.2013.29.

(14) Shen, Y.; Kan, D.; Tan, Z.; Wakabayashi, Y.; Shimakawa, Y. Tuning of Ferrimagnetism and Perpendicular Magnetic Anisotropy in $NiCo_2O_4$ Epitaxial Films by the Cation Distribution. *Phys Rev B* **2020**, *101* (9), 094412. https://doi.org/10.1103/PhysRevB.101.094412.

(15) Li, J.; Comstock, A. H.; Sun, D.; Xu, X. Comprehensive Demonstration of Spin Hall Hanle Effects in Epitaxial Pt Thin Films. *Phys Rev B* **2022**, *106* (18), 184420. https://doi.org/10.1103/PhysRevB.106.184420.

(16) Ludbrook, B. M.; Dubuis, G.; Puichaud, A.-H.; Ruck, B. J.; Granville, S. Nucleation and Annihilation of Skyrmions in Mn2CoAl Observed through the Topological Hall Effect. *Sci Rep* **2017**, *7* (1), 13620. https://doi.org/10.1038/s41598-017-13211-8.

(17) Chen, R. Y.; Zhang, R. Q.; Zhou, Y. J.; Bai, H.; Pan, F.; Song, C. Magnetic Field Direction Dependence of Topological Hall Effect like Features in Synthetic Ferromagnetic and Antiferromagnetic Multilayers. *Appl Phys Lett* **2020**, *116* (24), 242403. https://doi.org/10.1063/5.0011581.

(18) Vistoli, L.; Wang, W.; Sander, A.; Zhu, Q.; Casals, B.; Cichelero, R.; Barthélémy, A.; Fusil, S.; Herranz, G.; Valencia, S.; Abrudan, R.; Weschke, E.; Nakazawa, K.; Kohno, H.; Santamaria, J.; Wu, W.; Garcia, V.; Bibes, M. Giant Topological Hall Effect in Correlated Oxide Thin Films. *Nat Phys* **2019**, *15* (1), 67–72. https://doi.org/10.1038/s41567-018-0307-5.

(19) Liou, S.-H. Advanced Magnetic Force Microscopy Tips for Imaging Domains. In *Handbook of Advanced Magnetic Materials*; Springer US: Boston, MA, 2006; pp 374–396. https://doi.org/10.1007/1-4020-7984-2_10.

(20) Laraoui, A.; Ambal, K. Opportunities for Nitrogen-Vacancy-Assisted Magnetometry to Study Magnetism in 2D van Der Waals Magnets. *Appl Phys Lett* **2022**, *121* (6), 060502. https://doi.org/10.1063/5.0091931.

(21) Erickson, A.; Abbas Shah, S. Q.; Mahmood, A.; Fescenko, I.; Timalsina, R.; Binek, C.; Laraoui, A. Nanoscale Imaging of Antiferromagnetic Domains in Epitaxial Films of Cr2O3via Scanning Diamond Magnetic Probe Microscopy. *RSC Adv* **2022**, *13* (1), 178–185. https://doi.org/10.1039/d2ra06440e.

(22) Erickson, A.; Zhang, Q.; Vakili, H.; Li, C.; Sarin, S.; Lamichhane, S.; Jia, L.; Fescenko, I.; Schwartz, E.; Liou, S.-H.; Shield, J. E.; Chai, G.; Kovalev, A. A.; Chen, J.; Laraoui, A. Room Temperature Magnetic Skyrmions in Gradient-Composition Engineered CoPt Single Layers. *ACS Nano* **2024**, *18* (45), 31261–31273. https://doi.org/10.1021/acsnano.4c10145.

(23) Maletinsky, P.; Hong, S.; Grinolds, M. S.; Hausmann, B.; Lukin, M. D.; Walsworth, R. L.; Loncar, M.; Yacoby, A. A Robust Scanning Diamond Sensor for Nanoscale Imaging with Single Nitrogen-Vacancy Centres. *Nat Nanotechnol* **2012**, *7* (5), 320–324. https://doi.org/10.1038/nnano.2012.50.

(24) Vélez, S.; Schaab, J.; Wörnle, M. S.; Müller, M.; Gradauskaite, E.; Welter, P.; Gutgsell, C.; Nistor, C.; Degen, C. L.; Trassin, M.; Fiebig, M.; Gambardella, P. High-Speed Domain Wall Racetracks in a Magnetic Insulator. *Nat Commun* **2019**, *10* (1), 4750. https://doi.org/10.1038/s41467-019-12676-7.

(25) Avci, C. O.; Rosenberg, E.; Caretta, L.; Büttner, F.; Mann, M.; Marcus, C.; Bono, D.; Ross, C. A.; Beach, G. S. D. Interface-Driven Chiral Magnetism and Current-Driven Domain Walls in Insulating Magnetic Garnets. *Nature Nanotechnology*. Nature Publishing Group June 1, 2019, pp 561–566. https://doi.org/10.1038/s41565-019-0421-2.

(26) Ding, S.; Baldrati, L.; Ross, A.; Ren, Z.; Wu, R.; Becker, S.; Yang, J.; Jakob, G.; Brataas, A.; Kläui, M. Identifying the Origin of the Nonmonotonic Thickness Dependence of Spin-Orbit Torque and Interfacial Dzyaloshinskii-Moriya Interaction in a Ferrimagnetic Insulator Heterostructure. *Phys Rev B* **2020**, *102* (5), 054425. https://doi.org/10.1103/PhysRevB.102.054425.

(27) Caretta, L.; Rosenberg, E.; Büttner, F.; Fakhrul, T.; Gargiani, P.; Valvidares, M.; Chen, Z.; Reddy, P.; Muller, D. A.; Ross, C. A.; Beach, G. S. D. Interfacial Dzyaloshinskii-Moriya Interaction Arising from Rare-Earth Orbital Magnetism in Insulating Magnetic Oxides. *Nat Commun* **2020**, *11* (1), 1090. https://doi.org/10.1038/s41467-020-14924-7.

(28) Yang, H.; Thiaville, A.; Rohart, S.; Fert, A.; Chshiev, M. Anatomy of Dzyaloshinskii-Moriya Interaction at Co/Pt Interfaces. *Phys Rev Lett* **2015**, *115* (26), 267210. https://doi.org/10.1103/PhysRevLett.115.267210.

(29) Cao, A.; Zhang, X.; Koopmans, B.; Peng, S.; Zhang, Y.; Wang, Z.; Yan, S.; Yang, H.; Zhao, W. Tuning the Dzyaloshinskii–Moriya Interaction in Pt/Co/MgO Heterostructures through the MgO Thickness. *Nanoscale* **2018**, *10* (25), 12062–12067. https://doi.org/10.1039/C7NR08085A.

(30) Neubauer, A.; Pfleiderer, C.; Binz, B.; Rosch, A.; Ritz, R.; Niklowitz, P. G.; Böni, P. Topological Hall Effect in the A Phase of MnSi. *Phys Rev Lett* **2009**, *102* (18), 186602. https://doi.org/10.1103/PhysRevLett.102.186602.

(31) Demishev, S. V.; Glushkov, V. V.; Lobanova, I. I.; Anisimov, M. A.; Ivanov, V. Yu.; Ishchenko, T. V.; Karasev, M. S.; Samarin, N. A.; Sluchanko, N. E.; Zimin, V. M.; Semeno, A. V. Magnetic Phase Diagram of MnSi in the High-Field Region. *Phys Rev B* **2012**, *85* (4), 045131. https://doi.org/10.1103/PhysRevB.85.045131.

(32) Meng, K. K.; Zhao, X. P.; Liu, P. F.; Liu, Q.; Wu, Y.; Li, Z. P.; Chen, J. K.; Miao, J.; Xu, X. G.; Zhao, J. H.; Jiang, Y. Robust Emergence of a Topological Hall Effect in MnGa/Heavy Metal Bilayers. *Phys Rev B* **2018**, *97* (6), 060407. https://doi.org/10.1103/PhysRevB.97.060407.

(33) Ahmed, A. S.; Lee, A. J.; Bagués, N.; McCullian, B. A.; Thabt, A. M. A.; Perrine, A.; Wu, P. K.; Rowland, J. R.; Randeria, M.; Hammel, P. C.; McComb, D. W.; Yang, F. Spin-Hall Topological Hall Effect in Highly Tunable Pt/Ferrimagnetic-Insulator Bilayers. *Nano Lett* **2019**, *19* (8), 5683–5688. https://doi.org/10.1021/acs.nanolett.9b02265.

(34) Ohuchi, Y.; Matsuno, J.; Ogawa, N.; Kozuka, Y.; Uchida, M.; Tokura, Y.; Kawasaki, M. Electric-Field Control of Anomalous and Topological Hall Effects in Oxide Bilayer Thin Films. *Nat Commun* **2018**, *9* (1), 213. https://doi.org/10.1038/s41467-017-02629-3.

(35) Wang, W.; Daniels, M. W.; Liao, Z.; Zhao, Y.; Wang, J.; Koster, G.; Rijnders, G.; Chang, C. Z.; Xiao, D.; Wu, W. Spin Chirality Fluctuation in Two-Dimensional Ferromagnets with Perpendicular Magnetic Anisotropy. *Nat Mater* **2019**, *18* (10), 1054–1059. https://doi.org/10.1038/s41563-019-0454-9.

(36) Kar, U.; Singh, A. Kr.; Yang, S.; Lin, C.-Y.; Das, B.; Hsu, C.-H.; Lee, W.-L. High-Sensitivity of Initial SrO Growth on the Residual Resistivity in Epitaxial Thin Films of SrRuO3 on SrTiO3 (001). *Sci Rep* **2021**, *11* (1), 16070. https://doi.org/10.1038/s41598-021-95554-x.

(37) Denneulin, T.; Kovács, A.; Boltje, R.; Kiselev, N. S.; Dunin-Borkowski, R. E. Geometric Phase Analysis of Magnetic Skyrmion Lattices in Lorentz Transmission Electron Microscopy Images. *Sci Rep* **2024**, *14* (1). https://doi.org/10.1038/s41598-024-62873-8.

(38) Cho, J.; Kim, N.-H.; Lee, S.; Kim, J.-S.; Lavrijsen, R.; Solignac, A.; Yin, Y.; Han, D.-S.; van Hoof, N. J. J.; Swagten, H. J. M.; Koopmans, B.; You, C.-Y. Thickness Dependence of the Interfacial Dzyaloshinskii–Moriya Interaction in Inversion Symmetry Broken Systems. *Nat Commun* **2015**, *6* (1), 7635. https://doi.org/10.1038/ncomms8635.

(39) Chaurasiya, A. K.; Banerjee, C.; Pan, S.; Sahoo, S.; Choudhury, S.; Sinha, J.; Barman, A. Direct Observation of Interfacial Dzyaloshinskii-Moriya Interaction from Asymmetric

Spin-Wave Propagation in W/CoFeB/SiO2 Heterostructures Down to Sub-Nanometer CoFeB Thickness. *Sci Rep* **2016**, *6* (1), 32592. https://doi.org/10.1038/srep32592.

(40) Kan, D.; Xie, L.; Shimakawa, Y. Scaling of the Anomalous Hall Effect in Perpendicularly Magnetized Epitaxial Films of the Ferrimagnet NiCo2O4. *Phys Rev B* **2021**, *104* (13), 134407. https://doi.org/10.1103/PhysRevB.104.134407.

(41) Kimbell, G.; Sass, P. M.; Woltjes, B.; Ko, E. K.; Noh, T. W.; Wu, W.; Robinson, J. W. A. PHYSICAL REVIEW MATERIALS 4 , 054414 ( 2020 ) Two-Channel Anomalous Hall Effect in SrRuO 3. *Phys Rev Mater* **2020**, *4* (5), 54414. https://doi.org/10.1103/PhysRevMaterials.4.054414.

(42) Cheong, S.-W.; Fiebig, M.; Wu, W.; Chapon, L.; Kiryukhin, V. Seeing Is Believing: Visualization of Antiferromagnetic Domains. *NPJ Quantum Mater* **2020**, *5* (1), 3. https://doi.org/10.1038/s41535-019-0204-x.

(43) Doherty, M. W.; Manson, N. B.; Delaney, P.; Jelezko, F.; Wrachtrup, J.; Hollenberg, L. C. L. The Nitrogen-Vacancy Colour Centre in Diamond. *Phys Rep* **2013**, *528* (1), 1–45. https://doi.org/10.1016/j.physrep.2013.02.001.

(44) Casola, F.; van der Sar, T.; Yacoby, A. Probing Condensed Matter Physics with Magnetometry Based on Nitrogen-Vacancy Centres in Diamond. *Nat Rev Mater* **2018**, *3* (1), 17088. https://doi.org/10.1038/natrevmats.2017.88.

(45) Wang, X. S.; Yuan, H. Y.; Wang, X. R. A Theory on Skyrmion Size. *Commun Phys* **2018**, *1* (1), 31. https://doi.org/10.1038/s42005-018-0029-0.

(46) Balasubramanian, B.; Manchanda, P.; Pahari, R.; Chen, Z.; Zhang, W.; Valloppilly, S. R.; Li, X.; Sarella, A.; Yue, L.; Ullah, A.; Dev, P.; Muller, D. A.; Skomski, R.; Hadjipanayis, G. C.; Sellmyer, D. J. Chiral Magnetism and High-Temperature Skyrmions in B20-Ordered Co-Si. *Phys Rev Lett* **2020**, *124* (5), 057201. https://doi.org/10.1103/PhysRevLett.124.057201.

(47) Ullah, A.; Balamurugan, B.; Zhang, W.; Valloppilly, S.; Li, X.-Z.; Pahari, R.; Yue, L.-P.; Sokolov, A.; Sellmyer, D. J.; Skomski, R. Crystal Structure and Dzyaloshinski–Moriya Micromagnetics. *IEEE Trans Magn* **2019**, *55* (7), 7100305. https://doi.org/10.1109/TMAG.2018.2890028.

(48) Ullah, A.; Li, X.; Jin, Y.; Pahari, R.; Yue, L.; Xu, X.; Balasubramanian, B.; Sellmyer, D. J.; Skomski, R. Topological Phase Transitions and Berry-Phase Hysteresis in Exchange-Coupled Nanomagnets. *Phys Rev B* **2022**, *106* (13), 134430. https://doi.org/10.1103/PhysRevB.106.134430.

(49) Li, P.; Xia, C.; Zheng, D.; Wang, P.; Jin, C.; Bai, H. Observation of Large Low-Field Magnetoresistance in Spinel Cobaltite: A New Half-Metal. *Physica Status Solidi - Rapid Research Letters* **2016**, *10* (2), 190–196. https://doi.org/10.1002/pssr.201510402.

(50) Xiao, D.; Chang, M. C.; Niu, Q. Berry Phase Effects on Electronic Properties. *Rev Mod Phys* **2010**, *82* (3), 1959–2007. https://doi.org/10.1103/RevModPhys.82.1959.